\documentclass[11pt,onecolumn,a4paper]{article}
\usepackage{amssymb}

\usepackage[latin1]{inputenc}
\usepackage[T1]{fontenc}
\usepackage{cite}
\usepackage[dvips]{epsfig}
\usepackage[dvips]{color}
\usepackage{amsmath}
\usepackage{graphicx}
\usepackage{bm}
\usepackage{mathrsfs}
\usepackage{float}
\usepackage{slashed}
\usepackage[english]{babel}
\usepackage{subfigure}

\begin{document}

\title{\textbf{\Large Short wavelength electromagnetic propagation in
magnetized quantum plasmas}}
\author{J. Lundin, J. Zamanian, M. Marklund, and G. Brodin \\
\textit{Department of Physics, Ume\aa\ University, SE-901 87 Ume\aa, Sweden}}
\maketitle

\begin{abstract}
The quantum electrodynamical (QED) short wavelength correction on plasma wave propagation for a non-relativistic quantum plasma is investigated. A general dispersion relation for a thermal multi-component quantum plasma is derived. It is found that the classical dispersion relation for any wave mode can be modified to include quantum and short wavelength QED effects by simple substitutions of the thermal velocity and the plasma frequency. Furthermore, the dispersion relation has been modified to include QED effects of strong magnetic fields. It is found that strong magnetic fields together with the short wavelength QED correction will induce dispersion both in vacuum and in otherwise non-dispersive plasma modes. Applications to laboratory and astrophysical systems are discussed.
\end{abstract}

\section{Introduction}

There is currently a great deal of interest in quantum vacuum effects (see
e.g.\ Refs.\ \cite
{book,collectionA,collectionB,collectionC,collectionD,collectionE,collectionF,collectionG,collectionH,collectionI}
and references therein for an up-to-date selection) as well as in
non-relativistic quantum plasma systems \cite
{collection2A,collection2B,collection2C,collection2D,collection2E,collection2F,collection2G,collection2H,collection2I,collection2J,collection2K,collection2L,Haas-HarrisSheet}%
, starting from e.g.\ the Schr\"odinger model of the electron \cite
{Manfredi,holland} or the Pauli equation \cite{marklund-brodin,brodin-marklund}. 
Using a theory based on non-relativistic quantum mechanics, there have been studies of a wide range of plasma wave phenomena \cite{collection2C,collection2E,collection2F,collection2H} as well as astrophysical applications (e.g. Ref. \cite{collection2G}). In relativistic quantum theory, some of the literature has been motivated by the rapid progress in experimental techniques \cite{Marklund, exp1, exp2} suggesting experiments for different vacuum polarization effects,
\cite{collectionA, collectionB, collectionC, collectionF, ding-kaplan, Moulin1999, bernard, 
brodin-marklund-stenflo, eriksson-brodin-marklund-stenflo, shen-yu-wang, dipiazza, lundin}
(see also Ref.\ \cite{collectionH} for a discussion).  
For example, in Refs.\ \cite{collectionA,collectionF,ding-kaplan,Moulin1999,brodin-marklund-stenflo,eriksson-brodin-marklund-stenflo,shen-yu-wang,dipiazza,lundin} suggestions for experimental setups to verify photon-photon interaction are given, including the use of strong lasers (e.g. Ref.\ \cite{collectionA}) and dense plasma columns \cite{shen-yu-wang}. 
New experimental results have been presented in for example Ref.\ \cite{collectionB}, which reports a change in polarization of light traveling through a strong magnetic field (see also Refs.\ \cite{collectionC,collectionI} for further discussions). Other examples are Ref.\ \cite{burke} which reports electron-positron pair production due to a intense electromagnetic field interacting with high frequency photons, and Ref.\ \cite{bula} which investigates non-linear Compton scattering.
There are also studies of theoretical and/or observational nature 
\cite{collectionD, collectionE, adler, rozanov, soljacic-segev, marklund-brodin-stenflo, shukla-marklund-brodin-stenflo, marklund-shukla-brodin-stenflo ,stenflo-brodin-marklund-shukla, marklund-shukla-eliasson, marklund-shukla-stenflo-brodin-servin, marklund-brodin-stenflo-shukla, marklund-tskhakaya-shukla, marklund-eliasson-shukla, mendonca, brodin-marklund-eliasson-shukla}, which find applications in high intensity laser systems \cite{collectionD,collectionE,marklund-tskhakaya-shukla,marklund-eliasson-shukla} and in astrophysics \cite{adler, rozanov, soljacic-segev, marklund-brodin-stenflo,shukla-marklund-brodin-stenflo, marklund-shukla-brodin-stenflo ,stenflo-brodin-marklund-shukla, marklund-shukla-eliasson, marklund-shukla-stenflo-brodin-servin, marklund-brodin-stenflo-shukla,mendonca, brodin-marklund-eliasson-shukla}. 
For quantum plasma dynamics, a set of effective macroscopic field equations may be derived from the microscopic equations of motion for the relevant particles using a range of
techniques \cite{Manfredi,holland,Marklund}. Such effective models can be
very useful when trying to understand some of the many collective effects that may
occur in such systems, and can also give an indication on when one can
expect macroscopic phenomena to arise from the quantum realm. Astrophysical
environments exhibit many extremes \cite{Beskin-book,asseo}, e.g. magnetars 
\cite{magnetar}, and as such constitute interesting areas of applications
for quantum vacuum physics as well as relativistic quantum plasmas. For such applications
there have been a number of studies of relativistic quantum plasmas, e.g. Refs. \cite
{melrose,melrose-weise,baring-etal,harding-lai}. Indeed, in order to
properly investigate such extreme systems, a quantum field theoretical basis that incorporates
certain quantum vacuum effects is desirable.

Here, we will investigate the effects on non-relativistic quantum plasmas due to short wavelength corrections from quantum electrodynamics (QED). Such combined effects are indeed compatible, as the short wavelength corrections do not necessarily imply relativistic particle motions. In Section 2 we present the weak field short wavelength corrected Maxwell's equations and derive a quantum corrected fluid model of the plasma. These modified equations are used in Section 3 to derive a general dispersion relation for any plasma mode in a thermal multi-component plasma. The possibility of designing experiments where some of the above effects could be detected is of great fundamental interest. Such scenarios are discussed in Section 4. Furthermore, the quantum field theoretical correction due to strong magnetic fields, relevant to astrophysical environments, is discussed in Section 5 where the general dispersion relation is modified to also include such effects. In particular, we derive the dispersion relations for the possible plasma modes in environments similar to those in the vicinity of magnetars and pulsars. Finally, in Section 6 we summarize our results.

\section{Governing equations}

\subsection{Electromagnetic field equations}

The first order QED effects can effectively be modeled through the Heisenberg-Euler Lagrangian density \cite{Heisenberg-Euler, Schwinger}. This Lagrangian describes a vacuum perturbed by a slowly varying electromagnetic field. The effect of rapidly varying fields can be accounted for by adding a derivative correction to the Lagrangian \cite{Mamaev-1981}. This correction is referred to as the derivative QED correction or the short wavelength QED correction. Since the QED effect of electron-positron pair creation is not included in this model, the field strength must be lower than the critical field strength, $E_{\text{crit}}\equiv m_{e}^{2}c^{3}/\hbar e\approx 10^{16}\ \mathrm{V/cm^{-1}}$, and the frequency must remain lower than the Compton frequency $\omega _{e}\equiv m_{e}c^{2}/\hbar $ \cite{Marklund,probing}. Here $m_e$ is the electron mass, $c$ is the speed of light in vacuum, $e$ is the elementary charge and $\hbar$ is Plank's constant. The Heisenberg-Euler Lagrangian density with the derivative correction reads 
\begin{eqnarray}
&&\!\!\!\!\!\!\mathcal{L}=\mathcal{L}_{0}+\mathcal{L}_{HE}+\mathcal{L}_{D} 
\notag \\
&&\!\!\!\!=\frac{\varepsilon _{0}}{4}F_{ab}F^{ab}+\frac{\varepsilon
_{0}^{2}\kappa }{16}\left[ 4\left( F_{ab}F^{ab}\right) ^{2}+7\left( F_{ab}%
\widehat{F}^{ab}\right) ^{2}\right] +\sigma \varepsilon _{0}\left[ \left(
\partial _{a}F^{ab}\right) \left( \partial _{c}F_{\phantom{b} b}^{c}\right) -F_{ab}\Box
F^{ab}\right] , \nonumber \\  
\end{eqnarray}
where $\mathcal{L}_{0}$ is the classical Lagrangian density, while $\mathcal{L}_{HE}$ represents the Heisenberg-Euler correction due to first order strong field QED effects, $\mathcal{L}_{D}$ is the derivative correction, $\Box =\partial _{a}\partial ^{a}$ is the d'Alembertian, $F^{ab}$ is the electromagnetic field tensor and $\widehat{F}^{ab}=\epsilon^{abcd}F_{cd}/2$ where $\epsilon^{abcd}$ is the totally antisymmetric tensor. The parameter $\kappa =2\alpha ^{2}\hbar ^{3}/45m^{4}c^{5}$ gives the nonlinear coupling, $\sigma =(2/15)\alpha c^{2}/\omega _{e}^{2}$ is the coefficient of the derivative correction and $\alpha =e^{2}/4\pi \hbar c\varepsilon _{0}$ is the fine structure constant, where $\varepsilon _{0}$ is the free space permittivity. We obtain the field equations from the Euler-Lagrange equations $\partial _{b}\left[ \partial \mathcal{L}/\partial F_{ab}\right] =\mu _{0}j^{a}$ \cite{Rozanov-1998,Shukla-2004c}, 
\begin{equation}
\left( 1+2\sigma \Box \right) \partial _{a}F^{ab}=2\varepsilon _{0}\kappa
\partial _{a}\left[ \left( F_{cd}F^{cd}\right) F^{ab}+\tfrac{7}{4}\left(
F_{cd}\widehat{F}^{cd}\right) \widehat{F}^{ab}\right] +\mu _{0}j^{b},
\label{eq:general wave equation}
\end{equation}
where $j^{a}$ is the four-current and $\mu _{0}$ is the free space permeability.

For the remainder of this manuscript we will investigate low-amplitude linear wave perturbations in a magnetized quantum plasma. The strong field QED term proportional to $\kappa \left( F_{cd}F^{cd}\right) F^{ab}$ then contributes with a term that is quadratic in the unperturbed magnetic field $B_{0}$, and linear in the wave perturbation. However, due to the smallness of $\kappa $ such a term does not become dynamically important unless the unperturbed field strengths approaches $B_{0}\sim 10^{8}-10^{10}$ \textrm{T}, i.e. in pulsar or magnetar environments. An example of such effects will be considered in the end of section 5, but until then the strong field QED effects will be omitted. Following Ref.\ \cite{Lundin} we will keep the derivative QED corrections proportional to $\sigma $. This is a natural approach provided $cB_{0}/E_{\mathrm{crit}}\ll \omega /\omega _{e}$, where $\omega $ is a typical frequency of the field. The corresponding sourced Maxwell equations resulting from the derivative corrected field equation then become 
\begin{equation}
\left[ 1+2\sigma \left( -\frac{1}{c^{2}}\frac{\partial ^{2}}{%
\partial t^{2}}+\nabla ^{2}\right) \right] \nabla \cdot \mathbf{E}=\frac{\rho }{%
\varepsilon _{0}},
\end{equation}
\begin{equation}
\left[ 1+2\sigma \left( -\frac{1}{c^{2}}\frac{\partial ^{2}}{\partial t^{2}}%
+\nabla ^{2}\right) \right] \left(\nabla \times \mathbf{B} -\frac{1}{c^{2}}\frac{\partial \mathbf{E}%
}{\partial t}\right) =\mu _{0}\mathbf{j},
\label{eq:1}
\end{equation}
and the source free Maxwell equations are $\nabla \cdot \mathbf{B}=0$ and 
\begin{equation}
\nabla \times \mathbf{E}=-\frac{\partial \mathbf{B}}{\partial t}.
\label{eq:2}
\end{equation}
Here $\mathbf{j}=\sum_{s}q_{s}n_{s}\mathbf{u}_{s}$ is the current density, $\rho =\sum_{s}q_{s}n_{s}$ is the charge density, the index $s$ denotes the species of particles considered, $q_{s}$ is their charge, $n_{s}$ is the particle density and $\mathbf{u}_{s}$ is the particle velocity.

\subsection{Particle equations of motion}
The plasma can be described by a model which captures some of the quantum properties of the plasma particles.
We here follow Ref.\ \cite{Manfredi}, where the
particles are described by the statistical mixture of $N$ states $\psi_i$, $%
i=1,2,\dots,N$. The index $i$ sums over all particles independent of
species. The one-particle states satisfy the Schrödinger equation, which for
each state reads 
\begin{equation}  \label{shrod}
i \hbar \frac{\partial \psi_i}{\partial t} = -\frac{\hbar^2}{2m_i} \nabla^2
\psi_i + q_i\phi \psi_i,
\end{equation}
where $m_i$ and $q_i$ are the mass and the charge of the particle
respectively. The equations are coupled by Poisson's equation 
\begin{equation}  \label{poissons}
\nabla^2 \phi = -\frac{1}{\varepsilon_0}\sum_{i=1}^N q_ip_i|\psi_i|^2 
\end{equation}
where $p_i$ is the occupation probability of state $\psi_i$. This model
amounts to assume that all entanglement between particles are neglected. To
derive a fluid description we make the ansatz $\psi_i = \sqrt{n_i} \exp(i
S_i/\hbar) $ where $n_i$ is the particle density, $S_i$ is real, and the velocity of the 
$i$'th particle is $\textbf{u}_i = \nabla S_i / m_i$.
Next we define the global density and velocity for particles of species $s$ as $n_s
= \sum_{j} p_j n_j$ and $\mathbf{u}_s =\sum_{j} p_j n_j\mathbf{u_j}/n_s$,
where $j$ runs over all particles of species $s$. Inserting this into
equations \eqref{shrod} and \eqref{poissons}, and separating the real and the imaginary part, we obtain the continuity equation 
\begin{eqnarray}  \label{eq:3}
\frac{\partial n_s}{\partial t}+\nabla\cdot\left(n_s\mathbf{u}_s\right)=0,
\end{eqnarray}
and the momentum equation
\begin{eqnarray}  \label{eq:4a}
\frac{\partial \mathbf{u}_s}{\partial t}+\left(\mathbf{u}_s\cdot\nabla\right)%
\mathbf{u}_s=\frac{q_s}{m_s}\mathbf{E} 
-\frac{1}{m_sn_s}\nabla P+\frac{\hbar^2}{2m_s^2}\nabla\left(\frac{1}{%
\sqrt{n_s}}\nabla^2\sqrt{n_s}\right).
\end{eqnarray}
Here we have made the substitution $\sum_{j} p_j( \nabla^2 \sqrt{n_j} )/ 
\sqrt{n_j} \rightarrow (\nabla^2 \sqrt{n_s})/\sqrt{n_s}$, which is valid for
length scales larger than the Fermi length $\lambda_{Fs} \equiv
v_{Fs}/\omega_{ps}$, where $v_{Fs} \equiv
\hbar\left(3\pi^2\varepsilon_0\omega_{ps}^2/m_s^2q_s^2\right)^{1/3}$ and $\omega_{ps}\equiv \sqrt{%
q_s^{2}n_{0s}/\varepsilon _{0}m_s}$ is the Fermi velocity and the plasma frequency of species $s$. We stress that the pressure term contains both the fermion pressure, $P_{Fs}$, and the thermal pressure, $P_{t}$.

This model can be extended to include non-zero magnetic fields, by starting with a Hamiltonian which includes the vector potential. In the resulting model \cite{collection2D}, the continuity equation \eqref{eq:3} is unchanged, and the momentum equation \eqref{eq:4a} becomes
\begin{eqnarray}  \label{eq:4}
\frac{\partial \mathbf{u}_s}{\partial t}+\left(\mathbf{u}_s\cdot\nabla\right)%
\mathbf{u}_s=\frac{q_s}{m_s}\left(\mathbf{E}+\mathbf{u}_s\times\mathbf{B}\right)
-\frac{1}{m_sn_s}\nabla P+\frac{\hbar^2}{2m_s^2}\nabla\left(\frac{1}{%
\sqrt{n_s}}\nabla^2\sqrt{n_s}\right).
\end{eqnarray}

\section{The general dispersion relation}

Our next aim is to study the linear modes of the system described by eqs.\ (\ref{eq:1}), (\ref{eq:2}), (\ref{eq:3}) and (\ref{eq:4}). We linearize and Fourier decompose these equations and let $n_{0s}$ represent a constant unperturbed particle density while $n_{1s}$ denotes a density perturbation. Furthermore, the unperturbed velocity is put to zero and from now on $\textbf{u}_s$ denotes the wave perturbed velocity. For long wavelengths, such that $\hbar k/m\ll v_{F}\ll \omega /k$, the pressure term in eq.\ (\ref{eq:4}) can be written as $(1/m_{s}n_{s})\nabla \left( P_{ts}+P_{Fs}\right) =(1/n_{0s})\left(v_{ts}^{2}+\frac{3}{5}v_{Fs}^{2}\right) \nabla n_{1s}$ \cite{Manfredi}, where we have neglected the thermal pressure that particles of different species exert on each other. Here, $k$ is the perturbation wave number, $\omega$ is the perturbation frequency and $v_t$ is the thermal velocity. The pressure term and quantum correction in eq.\ (\ref{eq:4}) can now be collected into an effective term with an effective wavenumber dependent velocity $V_{s}$, 
\begin{equation}
V_{s}^{2}\equiv 
v_{ts}^{2}+\frac{3}{5}v_{Fs}^{2}+\frac{\hbar ^{2}k^{2}}{4m_{s}^{2}}.
\end{equation}
Below, the collective thermal and quantum effects described by $V$ will sometimes be referred to as effective thermal effects. We define the $z$-axis to be along $\mathbf{B}_{0}$ and the $y$-axis to be orthogonal to $\mathbf{k}$, such that $\mathbf{B}_{0}=B_{0}\hat{\mathbf{z}}$ and $\mathbf{k}=k_{x}\hat{\mathbf{x}}+k_{z}\hat{\mathbf{z}} \equiv k_{\bot}\hat{\mathbf{x}}+k_{\Arrowvert}\hat{\mathbf{z}}$. Now use eq.\ (\ref{eq:3}) to express $n_{1s}$ in terms of $\mathbf{u}_{s}$, and insert this result into eq.\ (\ref{eq:4}). Next we rewrite eq.\ (\ref{eq:4}) in the form $\boldsymbol\sigma _{s}^{-1}\cdot\mathbf{u}_{s}=\mathbf{E}$, and find $\boldsymbol\sigma _{s}$ such that $\mathbf{u}_{s}=\boldsymbol\sigma _{s}\cdot\mathbf{E}$. Using this result together with eq.\ (\ref{eq:2}), we can then rewrite eq.\ (\ref{eq:1}) in the form $\mathbf D \cdot\mathbf{E}=\mathbf{0}$ where
\begin{subequations}
\begin{equation}\label{eq:newE}
\mathbf D \equiv \mathbf I + \left( 
\begin{array}{ccc}
-n_{\Arrowvert}^{2}& 
0 & 
n_{\bot }n_{\Arrowvert} \\ 
0 & 
-n^{2} 
& 0\\ 
n_{\bot }n_{\Arrowvert}&
0 &
-n_{\bot }^{2}
\end{array}
\right) + \sum_{s}\boldsymbol{\chi}_s,
\end{equation}
and 
\begin{equation}\label{eq:suseptiblitiy}
\boldsymbol\chi_s =\left( 
\begin{array}{ccc}
-\frac{\omega _{ps}^{2}(\omega^{2}-k_{\Arrowvert}^{2}V_{s}^{2})}{\left( 1-\zeta \right) \omega _{V_{s}}^{4}} & 
-i\frac{\omega _{ps}^{2}\omega _{cs}(\omega ^{2}-k_{\Arrowvert}^{2}V_{s}^{2})}{\left( 1-\zeta \right) \omega \omega _{V_{s}}^{4}} & 
-\frac{\omega _{ps}^{2}k_{\bot}k_{\Arrowvert}V_{s}^{2}}{\left( 1-\zeta \right)\omega _{V_{s}}^{4}} \\ 
&  &  \\ 
i\frac{\omega _{ps}^{2}\omega _{cs}(\omega ^{2}-k_{\Arrowvert}^{2}V_{s}^{2})}{\left( 1-\zeta \right) \omega \omega _{V_{s}}^{4}} & 
-\frac{\omega _{ps}^{2}(\omega ^{2}-k^{2}V_{s}^{2})}{\left( 1-\zeta \right) \omega_{V_{s}}^{4}} & i\frac{\omega _{ps}^{2}\omega_{cs}k_{\bot}k_{\Arrowvert}V_{s}^{2}}{\left( 1-\zeta \right) \omega \omega _{V_{s}}^{4}} \\ 
&  &  \\ 
-\frac{\omega _{ps}^{2}k_{\bot}k_{\Arrowvert}V_{s}^{2}}{\left( 1-\zeta \right) \omega _{V_{s}}^{4}}&
 -i\frac{\omega_{ps}^{2}\omega _{cs}k_{\bot}k_{\Arrowvert}V_{s}^{2}}{\left( 1-\zeta \right) \omega\omega _{V_{s}}^{4}} &
-\frac{\omega _{ps}^{2}(\omega^{2}-k_{\bot}^{2}V_{s}^{2}-\omega _{cs}^{2})}{\left( 1-\zeta \right) \omega_{V_{s}}^{4}}\end{array}
\right).
\end{equation}
\end{subequations}
Here $\mathbf I$ is the identity matrix, $\sum_s \boldsymbol{\chi}_s$ is the short wavelength QED and quantum corrected plasma susceptiblitiy, $\omega _{cs}\equiv q_{s}B_{0}/m_{s}$ is the gyrofrequency, 
\begin{equation}
\zeta \equiv 2\sigma \left( \frac{\omega ^{2}}{c^{2}}-k^{2}\right) =-\frac{%
2\sigma \omega ^{2}}{c^{2}}\left( n^{2}-1\right) ,
\end{equation}
gives the derivative QED correction due to rapidly varying fields, $n_{\bot }\equiv k_{\bot}c/\omega $ and $ n_{\Arrowvert}\equiv k_{\Arrowvert}c/\omega$ is the index of refraction orthogonal respectively parallel to the magnetic field such that $n=\sqrt{n_{\bot }^{2}+n_{\Arrowvert}^{2}}$ is the total index of refraction, and we define
\begin{equation}
\omega _{V_{s}}^{4}\equiv \left( \omega ^{2}-k_{\bot}^{2}V_{s}^{2}\right)
\left( \omega ^{2}-k_{\Arrowvert}^{2}V_{s}^{2}\right) -\omega _{cs}^{2}\left( \omega
^{2}-k_{\Arrowvert}^{2}V_{s}^{2}\right) -k_{\bot}^{2}k_{\Arrowvert}^{2}V_{s}^{4}.
\end{equation}
Setting the determinant of $\mathbf D$ equal to zero gives the dispersion relation for all plasma modes. From expression (\ref{eq:suseptiblitiy}) it becomes transparent that the dispersion relation for any classical plasma mode can be modified to include the quantum correction by making the substitution
\begin{subequations}
\begin{equation}\label{eq:subst1}
v_{ts}\rightarrow V_{s},
\end{equation}
and the short wavelength QED contribution can be included by making the substitution
\begin{equation}\label{eq:subst2}
\omega _{ps}^{2}\rightarrow \frac{\omega _{ps}^{2}}{1-\zeta }.
\end{equation}
\end{subequations}
Note that the substitution of $\omega _{ps}$ should \emph{not} be made within the expression for $v_{Fs}$.

In order to demonstrate the usefulness of our results (\ref{eq:subst1}) and (\ref{eq:subst2}) and verify their applicability, we compare with previous results. As is well known, the classical dispersion relation for ion-acoustic waves with zero ion-temperature reads $\omega/k = v_{te}/\sqrt{1+k^2v_{te}^2m_e/m_i}$, where indices $e$ and $i$ denotes electron and ion, respectively. Making the substitution (\ref{eq:subst1}) in the classical expression immediately give the dispersion relation (23) of Ref. \cite{collection2C}. Furthermore, the classical dispersion relation for almost perpendicular propagating electrostatic ion-cyclotron waves reads $\omega^2=\omega_{ci}^2+k_{\bot}^2 v_{te}^2m_e/m_i$. Again, making the substitution (\ref{eq:subst1}) in the classical formula we obtain agreement with the dispersion relation (13) in Ref. \cite{collection2L} in the appropriate limit. 

Moreover, the classical zero temperature dispersion relations for plasma oscillations $\omega^2=\omega_p^2$, electromagnetic waves in an unmagnetized plasma $\omega^2=k^2c^2+\omega_p^2$, electromagnetic waves propagating parallel to an external magnetic field $(\omega \pm \omega_c)(\omega-k^2c^2/\omega)-\omega_p^2=0$, and extraordinary waves $(\omega_p^2-\omega^2)(\omega_p^2+k^2c^2-\omega^2)+\omega_c^2(k^2c^2-\omega^2)=0$, all give agreement with corresponding relations in Ref. \cite{Lundin}, when substitution (\ref{eq:subst2}) in the classical expressions.

\section{Laboratory application}
It would be of great fundamental interest if the quantum or derivative QED effects could be detected in experiments. Firstly, we will be concerned with the possibility of detecting the derivative QED contribution in the laboratory. To single out the small dispersive QED effects from the general dispersive effects of a plasma requires detailed knowledge of the plasma parameters for an experiment to be conclusive. We now investigate whether it may be possible to detect the short wavelength QED effects by studying how the index of refraction scales with frequency close to a plasma resonance, rather than trying to determine the actual phase shift caused by the derivative QED correction. For this purpose, assume that we let several laser beams with slightly different frequencies pass close to each other through the same plasma column in such a way that all beams experience the same plasma density perturbations. In this way, precise knowledge of the plasma density profile is not necessary for interpreting the results. The phase shift due to the difference in optical path length between a beam passing through the plasma column and one that does not pass through the plasma column can be measured with high precision \cite{interferometry}. How the index of refraction scales with the frequency can be estimated by comparing the phase shifts of the different laser beams.

\begin{figure}
\begin{minipage}{0.65\linewidth}
\includegraphics[width=1\textwidth]{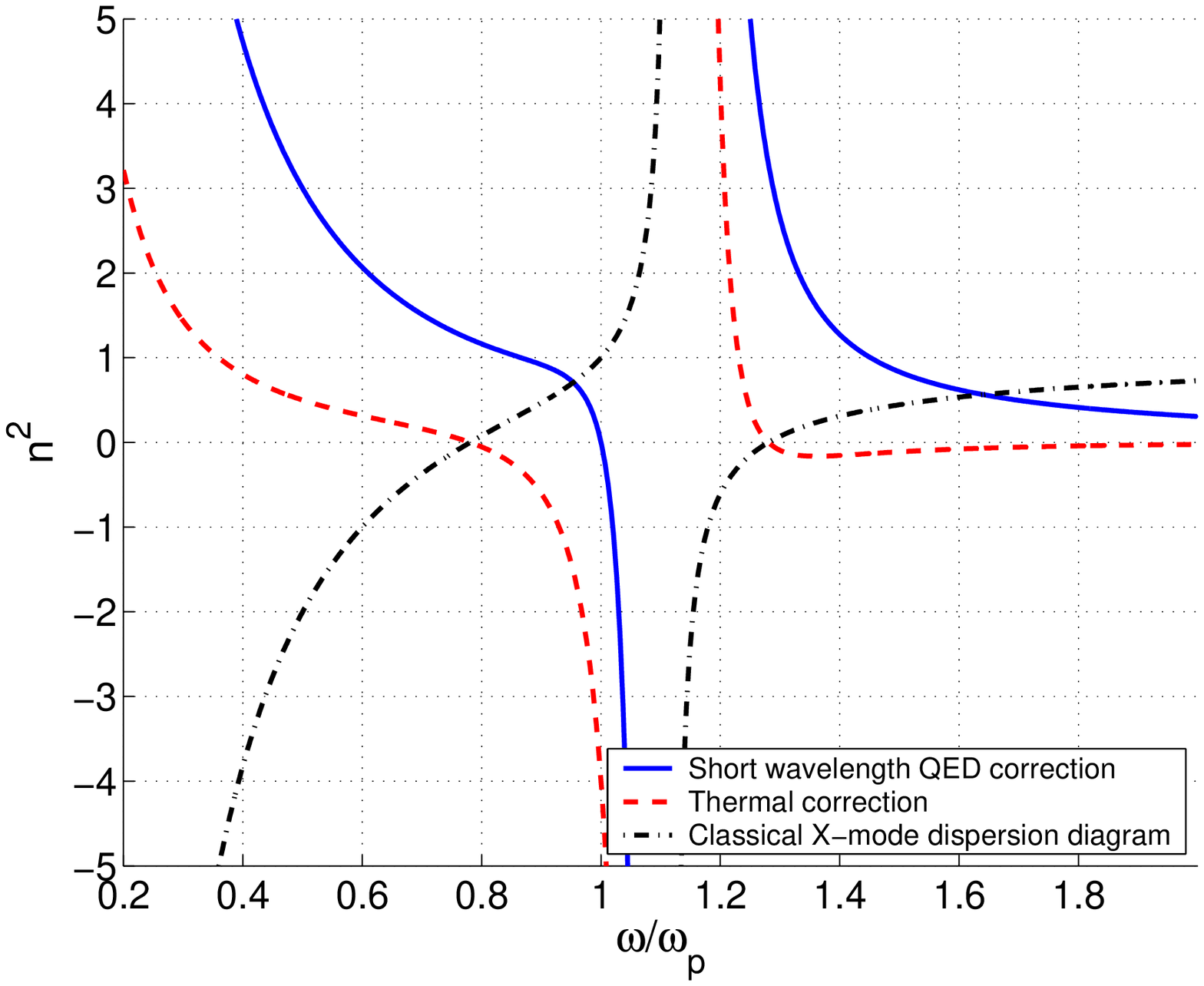} 
\begin{footnotesize}
\begin{center}
\textbf{(a)}
\end{center}
\end{footnotesize}
\end{minipage}
\begin{minipage}{0.35\linewidth}
\includegraphics[width=1\textwidth]{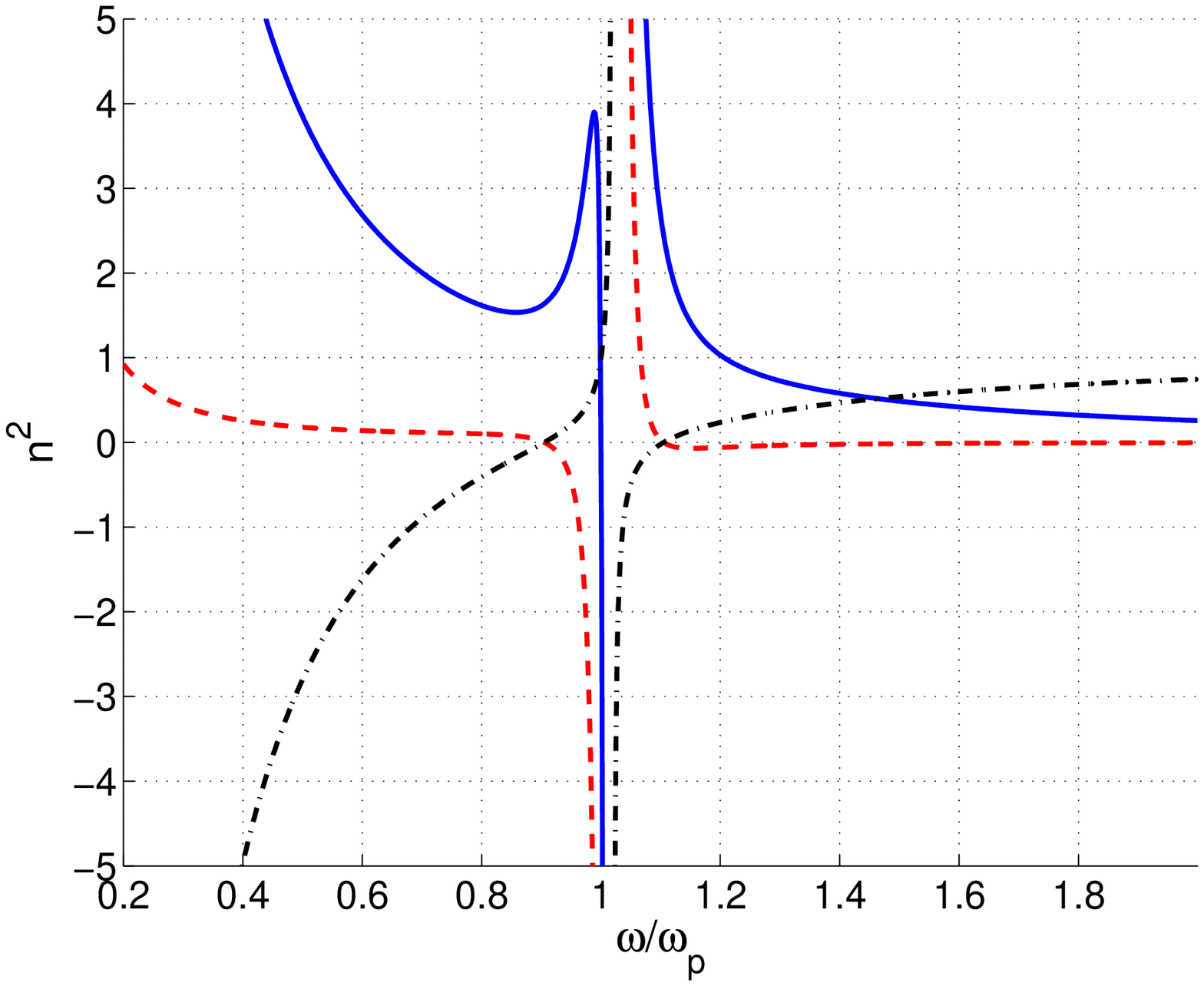} 
\begin{footnotesize}
\begin{center}
\textbf{(b)}
\end{center}
\end{footnotesize}
\includegraphics[width=1\textwidth]{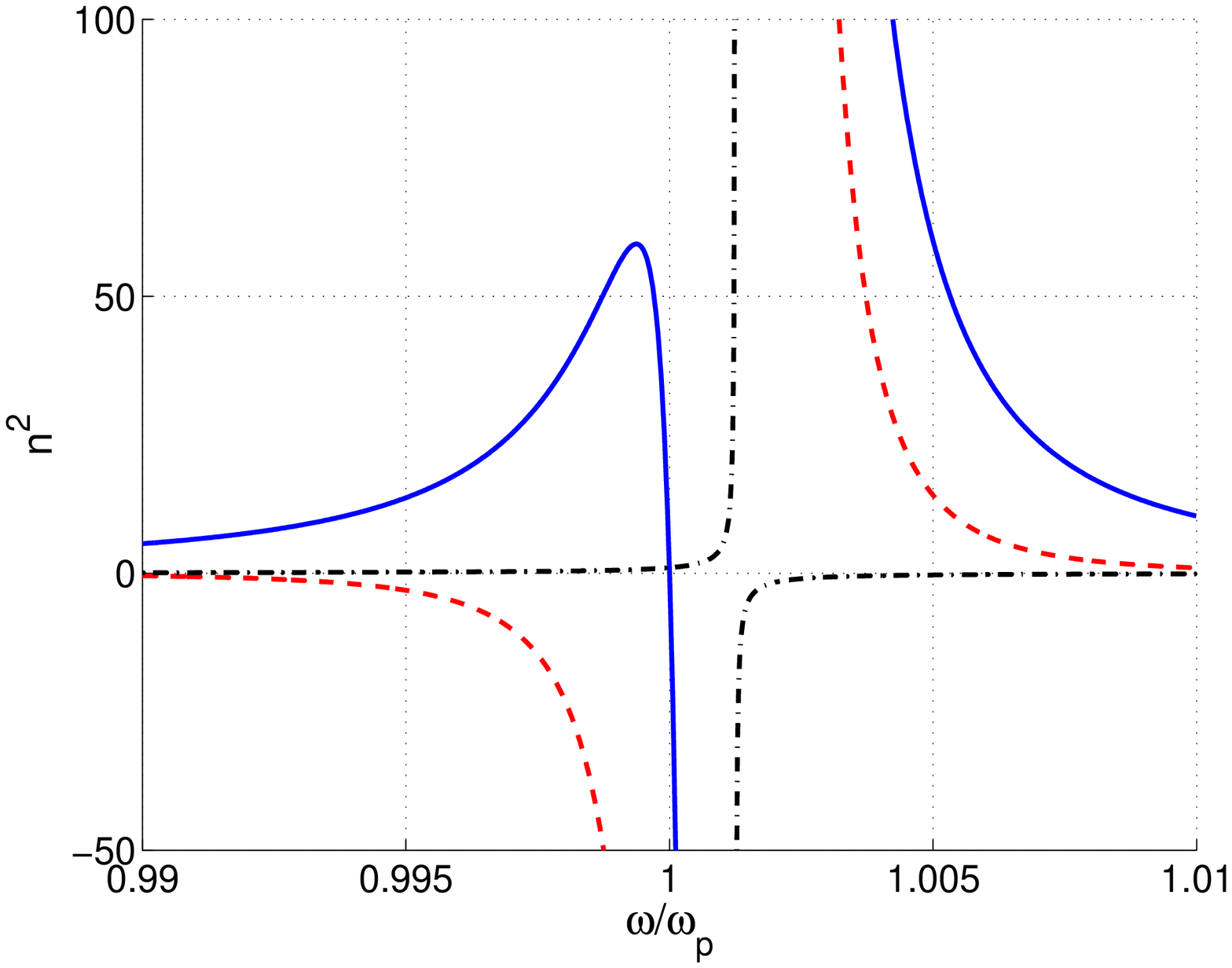} 
\begin{footnotesize}
\begin{center}
\textbf{(c)}
\end{center}
\end{footnotesize}
\end{minipage}
\caption{The derivative QED correction (dashed line) and the thermal and quantum correction (solid line) to the index of refraction, $n$, for an $X$-wave (dash-dot line). Here, $\bar{\protect\sigma}$ and $\bar V$ are both normalized to one and the magnetic field strength is chosen so that $\bar{\protect\omega}_{c}=0.5$ in Fig.\ $\left.1a\right)$, $\bar{\protect\omega}_{c}=0.2$ in Fig.\ $\left.1b\right)$ and $\bar{\protect\omega}_{c}=0.05$ in
Fig.\ $\left.1c\right)$}
\label{fig:Compare}
\end{figure}

Since the short wavelength QED effect is enhanced for large wavenumbers, see Ref.\ \cite{Lundin}, one of the most interesting modes when considering the derivative QED correction is the extraordinary mode ($X$-wave). The classical dispersion relation for an $X$-wave in a thermal electron plasma reads
\begin{eqnarray}\label{eq:classicalXmode}	n^2=1-\frac{\omega_p^2}{\omega^2}\frac{\left(\omega^2-k^2v_t^2-\omega_p^2\right)}{\left(\omega^2-k^2v_t^2-\omega_c^2-\omega_p^2\right)}.
\end{eqnarray}
We include the quantum and the derivative QED contributions, by making the substitutions given in (\ref{eq:subst1}) and (\ref{eq:subst2}). The index of refraction to the lowest non vanishing order in $\sigma$ and $V$ then becomes
\begin{eqnarray}  \label{eq:expansion}
n^2 &\approx& 1-\frac{1}{\bar{\omega}^2}\frac{\left(\bar{\omega}^2-1\right)}{%
\left(\bar{\omega}^2-\bar{\omega}_{c}^2 -1\right)}+ 
\frac{\bar \sigma}{\bar \omega^2}\left(\frac{\left(\bar \omega^2-1\right)\left(\bar \omega^2-2\right)}{\left(\bar \omega^2-\bar \omega^2_c-1\right)^2} + \frac{\left(\bar \omega^2-1\right)^2}{\left(\bar \omega^2-\bar \omega^2_c-1\right)^3} \right) + \notag \\ 
&+&\frac{\bar V^2}{\bar\omega^2}\left( \frac{\bar\omega^2}{%
\left(\bar\omega^2-\bar\omega_{c}^2-1\right)} - \frac{\bar\omega^4-1}{%
\left(\bar\omega^2-\bar\omega_{c}^2-1\right)^2} + \frac{\left(\bar\omega^2-1%
\right)^2}{\left(\bar\omega^2-\bar\omega_{c}^2-1\right)^3} \right) + O(\bar{%
\sigma}^2,\bar V^4),
\end{eqnarray}
where the parameters have been normalized according to 
\begin{equation}\label{eq:subst}
\bar{\sigma }\equiv\frac{2\omega _{p}^{2}}{c^{2}}\sigma, \qquad \bar V \equiv \frac{V}{c}, \qquad \bar{\omega }%
\equiv\frac{\omega }{\omega _{p}} \qquad \text{and} \qquad \bar{\omega}_{c}
\equiv \frac{\omega_{c}}{\omega_p}.
\end{equation}
Here we have disregarded the $k$-dependence of the effective thermal velocity and treated $V$ as a constant. The classical dispersion diagram for an $X$-wave has been illustrated in Fig.\ (\ref{fig:Compare}). The derivative QED contribution and the effective thermal contribution given in eq.\ (\ref{eq:expansion}) has been illustrated separately with $\bar{\sigma}$ and $\bar V$ normalized to one.

For an $X$-wave close to plasma resonance, eq.\ (\ref{eq:expansion}) shows that the leading term in the derivative QED contribution to the index of refraction has a cubic dependence on the resonance term, $\left(\bar \omega^2-\bar \omega^2_c-1\right)^{-1}$, in contrast to the classical contribution which is linear in this term. For detection to be conclusive, the thermal effects must be much smaller than the derivative QED effects in the regime of detection. Unfortunately, the leading term for the thermal contribution has the same form as that of the derivative QED contribution. This gives us the condition $ V < \omega_p\sqrt{2\sigma}$, 
which must be fulfilled for the thermal effects to be smaller than short wavelength QED effects close to a plasma resonance. Even if the plasma temperature is low, the Fermi velocity ($\propto \omega_p^{2/3}$) will always remain greater than $\omega_p\sqrt{2\sigma}$ for laboratory plasma densities. Thus, derivative QED effects will be negligible compared to the effective thermal effects in all laboratory plasmas. 

Even if we allow for higher densities, we can still expect the short wavelength QED contribution to be negligible. Moreover, care has to be taken since high plasma densities may imply a relativistic Fermi velocity, while we in our model have assumed a non-relativistic plasma. In contrast to being close to resonance, we next consider a regime far from resonance. In particular, we require $\bar \omega \ll 1$. For $\omega$ to be greater than the cutoff frequency while the condition $\bar \omega \ll 1$ is satisfied, we must require $\bar \omega_c \gg 1$. The leading term of the thermal contribution and the derivative QED correction in (\ref{eq:expansion}) approximately becomes $\bar V^2 /\bar \omega_c^4 \bar \omega^2$ and $4\bar \sigma/\bar \omega_c^4 \bar \omega^2$, respectively. Hence, the condition under which the short wavelength QED contribution is greater than the effective thermal contribution becomes approximately $V < \omega_p \sqrt{4\sigma}$. Thus, derivative QED effects will be negligible compared to effective thermal effects also in this regime.

\begin{figure}
\includegraphics[width=0.7\columnwidth]{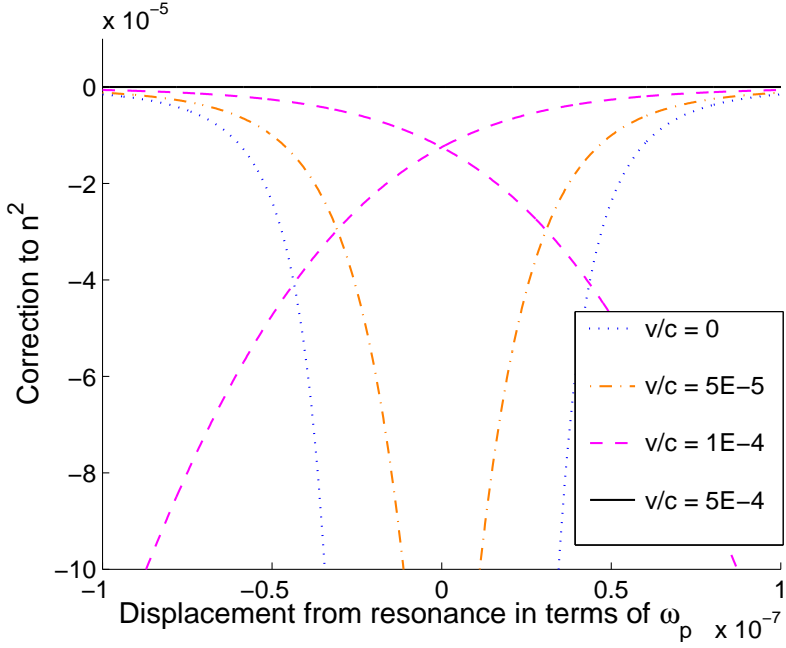}
\caption{The quantum correction to the index of refraction for an $X$-wave at different values of $\bar v$. The plasma frequency is in the infrared regime ($\protect\omega_p=10^{14}\ \mathrm{Hz}$) and the magnetic field strength is $1\ \mathrm{Tesla}$ ($\bar \omega_c = 0.001$).}
\label{fig:Thermal}
\end{figure}

One could in principle try to use the same technique outlined above to detect the dispersive effects due to the quantum correction. With the quantum correction included, the effective velocity $V^2$, should be written as 
\begin{equation}
V^2 = v^2 + n^2\frac{\hbar^2\omega^2}{4m^2c^2},
\end{equation}
where $v^2 = \frac{3}{5}v_F^2+v_t^2$. In order to get a phase shift due to the quantum contribution of approximately $\pi$ in a 10 m long plasma column with a plasma frequency in the infrared regime ($\omega_p \approx 10^{14}\ \mathrm{Hz}$), $\bar v = v/c$ must be smaller than $10^{-7}$ for the quantum correction to be the leading effect. The Fermi velocity alone gives a much greater contribution at these densities, $v_F/c\approx 10^{-4}$. Furthermore, the thermal effect removes the resonance in the dispersion diagram at high temperatures. In this regime, the quantum correction consequently stretches across the resonance as can be seen in Fig.\ (\ref{fig:Thermal}).

\section{Strong field QED effects}
Another interesting regime is that of extreme magnetic fields. In the vicinity of magnetars and pulsars, the magnetic field $B_{0}$ can reach $10^{10}-10^{11}\ \mathrm{T}$ close to the surface. Returning to the full wave equation (\ref{eq:general wave equation}) accounting for strong field effects, we note that the term proportional to $\kappa $ contributes with terms that are linear in the wave field and quadratic in $B_{0}$. Following Ref.\ \cite{Brodin} we find that the tensor in expression (\ref{eq:newE}) can be modified to include the QED effect of strong magnetic fields by adding the correction 
\begin{equation}
\boldsymbol\chi _{\text{{\tiny {QED}}}}=-4\frac{\xi }{1-\zeta }\left( 
\begin{array}{ccc}
1-n_{\Arrowvert}^{2} & 0 & n_{\bot }n_{\Arrowvert} \\ 
0 & 1-n^{2}-2n_{\bot } & 0 \\ 
n_{\bot }n_{\Arrowvert} & 0 & -\tfrac{5}{2}-n_{\bot }^{2}\label{eq:QEDcorrection-tensor}
\end{array}
\right) ,
\end{equation}
where $\xi \equiv \kappa \varepsilon _{0}c^{2}B_{0}^{2}=(\alpha /90\pi )(cB_{0}/E_{\text{crit}})^{2}$. In magnetar environments with extreme magnetic fields we expect the presence of a electron-positron plasma \cite{Beskin-book}. Assuming $n_{e0}\approx n_{p0}$ as well as $\omega _{p},\omega \ll \omega _{c}$, the $(1\!:\!2)$,$(2\!:\!1)$,$(2\!:\!3)$ and $(3\!:\!2)$ elements of expression (\ref{eq:suseptiblitiy}) are small and can be neglected. The dispersion relation for a plasma in a strong magnetic field where strong field QED and short wavelength QED corrections are included can now be derived from expression (\ref{eq:newE}) with expression\ (\ref{eq:QEDcorrection-tensor}) added. We find that the dispersion relation separates into two modes, 
\begin{subequations}
\begin{equation}
\left( 1-n^{2}\right) \left( 1-\zeta -4\xi \right) -8\xi n_{\bot }^{2}+\frac{%
\omega _{p}^{2}\left( \omega ^{2}-k^{2}V^{2}\right) }{\omega _{c}^{2}\left(
\omega ^{2}-k_{\Arrowvert}^{2}V^{2}\right) }\approx 0  \label{eq:strong-mode-1}
\end{equation}
and 
\begin{equation}
\left( 1-n^{2}\right) \left( 1-\zeta -4\xi \right) -\left( -14\xi +\frac{%
\omega _{p}^{2}}{\left( \omega ^{2}-k_{\Arrowvert}^{2}V^{2}\right) }\right) \left(
1-n_{\Arrowvert}^{2}\right) \approx 0,  \label{eq:strong-mode-2}
\end{equation}
\end{subequations}
where $\omega _{p}^{2}=\omega _{p,e}^{2}+\omega _{p,p}^{2}$. Terms proportional to $\omega _{p}^{2}/\omega _{c}^{2}$ in the second mode has been dropped, and $\xi \ll 1$ is assumed. Eqs.\ (\ref{eq:strong-mode-1}) and (\ref{eq:strong-mode-2}) generalized the dispersion relations derived in Ref.\ \cite{Brodin} to account for quantum and thermal effects, as well as QED derivative corrections. We note that if we let $\zeta \rightarrow 0$, the group velocity of the mode (\ref{eq:strong-mode-1}) becomes a constant in the zero temperature limit, whereas a nonzero $\zeta $ implies dispersive propagation. While the dispersive effects are small for most part of the spectrum, we note that for frequencies approaching the Compton frequency  $\omega _{e}$ together with field strengths approaching the critical field $E_{\text{crit}}$, dispersion may be appreciable. Strictly speaking, radiation with $\omega >\omega _{e}$ propagating in external fields with $cB_{0}>E_{\text{crit}}$ lies outside the validity of our model, and the dispersion relations (\ref{eq:strong-mode-1}) and (\ref{eq:strong-mode-2}) may thus not be entirely accurate in this regime. Nevertheless, our results here show convincingly that such electromagnetic wave propagation must be significantly dispersive. For high intensities, radiation in pulsar and magnetar environments with the above properties give raise to effects such as pair production and photon splitting (see e.g. Ref. \cite{harding-lai} and references therein). The competition between these two types of processes may determine the pair-plasma density \cite{Baring} in the vicinity of the pulsars and magnetars, and thereby have observable consequences. Normally, such effects are described without accounting for the dispersive wave properties \cite{adler,brodin-marklund-eliasson-shukla}, but we conclude here that our above results indicate that wave dispersion can play a role in this context.  

\section{Summary}

In this paper we have investigated how the derivative QED correction and the quantum correction affects plasma wave propagation in a multi-component thermal plasma. We have derived the generalized dispersion relation for all plasma modes. Specifically, it is found that the dispersion relation for any classical plasma mode can be modified to include the quantum contribution and the derivative QED correction by simple substitutions of the thermal velocity and the plasma frequency according to (\ref{eq:subst1}) and (\ref{eq:subst2}). Furthermore, the QED effect of a strong magnetic field has also been included in our plasma description.

We have also investigated the difficulties with detection of the short wavelength QED correction and the quantum correction. It is found that dispersive effects due to the Fermi velocity in plasma wave propagation will dominate in laboratory regimes. We conclude that the detection of the short wavelength QED correction or the quantum correction to the dispersion relation in laboratory plasma regimes is difficult. However, there may perhaps be ways around these particular problems.

Finally, we point out that there may exist astrophysical environments where strong field QED effects can be of importance. In particular, magnetars and pulsars offer environments with extreme magnetic fields. We have derived the dispersion relations for the two plasma modes likely to dominate in these environments. The short wavelength QED effect will give rise to dispersion in wavemode (\ref{eq:strong-mode-1}) which otherwise would be dispersionless in the zero temperature limit. Furthermore, strong magnetic fields will induce dispersive effects in vacuum. This is a unique property arising from the short wavelength QED correction. The dispersion can be significant for field strengths close to or above the critical field strength.\\
\\
\\
\\
\textbf{Acknowledgments}\\
\\
This research was supported by the Swedish Research Council Contract No.\ 621-2004-3217.\\

\pagebreak

\end{document}